\documentstyle[12pt]{article} 
  \newtheorem{df}{Definition}[section]
  
  \newtheorem{lem}[df]{Lemma}
  \newtheorem{rmk}[df]{Remark}
  \newtheorem{prop}[df]{Proposition}
  
  \newcommand{\qed}{\hbox{\rule{4pt}{8pt}}}
  \makeatletter
   
   \@addtoreset{equation}{section}
  \makeatother

\begin{document}
 \title{Born-Infeld Action in $(n+2)$-Dimension, \\
        the Field Equation \\ and a Soliton Solution}
 \author{Tatsuo SUZUKI\thanks{e-mail address : 695m5050@mn.waseda.ac.jp} \\
  Department of Mathematics \\ 
  Waseda University \\
  3-4-1 Okubo, Shinjuku-ku, \\
  Tokyo 169 \\
  Japan}
 \date{}
 \maketitle
\begin{abstract}
 In this paper we introduce the $(n+2)$-dimensional Born-Infeld action with a dual field strength $\tilde{H}$. We compute the field equation by using Schur polynomials and give a soliton solution. 
\end{abstract}
 \section{Introduction}

The Born-Infeld action was proposed by \cite{B-I} as a non-linear extension of the Maxwell theory. This theory has a dyon-like solution. Recently, the action of this type often appears in the context of M theory fivebrane. Moreover, duality of the action was studied by \cite{Tse}. Therefore, it is important to investigate and generalize this one. 

In \cite{P-S}, Perry and Schwarz constructed a theory in 6-dimensional Minkowski space whose action is given by 
\begin{eqnarray*}
 S_6 &=& \int d^6 x 
          \left( \frac 12 \tilde{H}^{\mu \nu } \partial_5 B_{\mu \nu }
                 +g(\tilde{H}) \right) \\
     && \quad \mbox{where} \quad g(\tilde{H})=
          2(\sqrt{-det(\eta +i\tilde{H})}-1) .
\end{eqnarray*}
Their theory has also a dyon-like solution as in the Born-Infeld theory, \cite{P-S}, \cite{Fuj}. A way to investigate this theory is to generalize this theory to any dimensional one. But we find that it is impossible to generalize in keeping the form of both the first term and the second term (Born-Infeld term) of $S_6$, because they use the self-duality in 6-dimensions. 

In consideration of this point, we shall in this paper generalize the Born-Infeld term of $S_6$ only to $(n+2)$-dimensional one. We calculate the field equation of the action and construct a 1-soliton solution of the equation. This solution corresponds to an extension of the Dirac monopole. In our theory, we don't use dimensional reduction as in \cite{P-S}, but a theory with dimensional reduction is straightforward. 

The contents of this paper are as follows. In section 2, we prepare notations and some mathematical tools. In particular, Schur polynomials play an important role to calculate the field equation. In section 3, we define the Born-Infeld action in $(n+2)$-dimension and calculate its field equation. This equation corresponds to the Yang-Mills equation of the Maxwell theory. To find solutions of the field equation, we need to consider the Bianchi identity together. In section 4, we reduce these field equations using an ansatz. The Bianchi identity and the Yang-Mills equation become the Laplace equation and a kind of an equation of motion respectively. In section 5, we construct 1-soliton solution using the ansatz above. We describe the solution using Cartesian coordinates and polar coordinates. In section 6, we summarize our theory and discuss $N$-soliton solutions. 

 \section{Preliminaries}

Let $(M^{n+2},\eta )$  be $(n+2)$-dimensional Minkowski spaces whose coordinate is 
$(x_0,\cdots,x_{n+1})$ , where 
$$ \eta =(-,\underbrace{+,\cdots,+}_{n+1}). $$

Let $B_{\rho_1 \cdots \rho_{n-1}}$ be a $(n-1)$-form valued gauge field and let 
$ H_{\rho_1 \cdots \rho_n}=\partial_{[ \rho_1} B_{\rho_2 \cdots \rho_n ] }$  be its field strength; $H=dB$.

We define the dual field of $H$ with respect to $\eta $ by
$$
 \tilde{H}^{\mu \nu }
   =\frac {1}{n!} \epsilon^{\mu \nu \rho_1 \cdots \rho_n} 
    H_{\rho_1 \cdots \rho_n}, $$
where
$$ 
   \left\{
    \begin{array}{l}
     \mu , \nu ,\cdots =0,1,2,\ldots ,n+1 \\
     \epsilon^{0 1 2 \cdots n+1}=1.
    \end{array}
   \right.
$$

As usual, we use the following matrix notation:
$$ \tilde{H}=(\tilde{H}_{\mu \nu }), \quad \eta=(\eta_{\mu \nu }), $$
$$ (\tilde{H}^{2k-1})_{\mu \nu }=\tilde{H}_{\mu \rho_1} \tilde{H}^{\rho_1 \rho_2}
     \cdots \tilde{H}_{\rho_{2k-2} \nu }. $$
These are $(n+2) \times (n+2)$ matrices.

 For $j=0,1,2,\ldots,m$, let $S_{1^j}(z)$ be Schur polynomials with variables $z=(z_1, \cdots ,z_m)$ corresponding to the partition $1^j=(\underbrace{1,\cdots,1}_{j})$, which are given by
\begin{equation}
   S_{1^j}(z)=\sum_{{\scriptstyle k_1+2k_2+ \cdots +mk_m=j}\atop
                    {\scriptstyle k_1 \geq 0,\cdots ,k_m \geq 0}}
              (-1)^{k_2+k_4+ \cdots +k_{2 \left[\frac m 2\right]}} 
              \frac1 {k_1! \cdots k_m!} z_1^{k_1} \cdots z_m^{k_m} 
 \label{eqn:Schur}
\end{equation}
(see, for example, \cite{Mac}).
\begin{rmk}
 By the change of variables
 \begin{equation}
   z_j=\frac 1j (\epsilon_1^j+ \cdots + \epsilon_m^j) 
    \quad \quad j=1, \cdots ,m, 
 \end{equation}
 $ S_{1^j}(z)$'s become the fundamental symmetric polynomials of degree $j$ with respect to $\epsilon_1, \cdots , \epsilon_m$.
\end{rmk}

We shall use the next lemmas:

\begin{lem}
 For $i=1,2,\ldots,m$, 
\begin{equation}
   \frac{\partial }{\partial z_i} \left(\sum_{j=0}^m S_{1^j}(z) \right)
     =(-1)^{i-1} \sum_{j=0}^{m-i} S_{1^j}(z).
\end{equation}
\end{lem}

\begin{lem}\label{lem:det} 
 For $n=2m-1$ or $2m-2$, put 
 $$
 z_j \equiv \frac{1}{2j}Tr(\tilde{H} \eta )^{2j} \quad (j=1,\cdots,m). 
 $$
 Then we have
\begin{equation}
 -det(\eta +i\tilde{H})=\sum_{j=0}^m S_{1^j}(z).
\end{equation}
\end{lem}

 \section{Action and the field equations}

 We define the Born-Infeld action in $(n+2)$-dimension by
\begin{equation}
  S_{n+2} = S_{n+2}(B) =\int d^{n+2}x
   \left\{ 2(\sqrt{-det(\eta +i\tilde{H})}-1) \right\}. 
\end{equation}
We calculate the variation of this action with respect to $B_{\rho_1 \cdots \rho_{n-1}}$. By the above lemmas we have, for $n=2m-1$ or $2m-2$,
\begin{eqnarray*}
 \delta S_{n+2} 
 &=& \int d^{n+2}x
  \frac{1}{\sqrt{\sum_{j=0}^m S_{1^j}(z)}} \sum_{k=1}^m 
   \frac{\partial }{\partial z_k} \left(\sum_{j=0}^m S_{1^j}(z) \right) \\
 && \times \left\{-\frac {1}{(n-1)!} \epsilon^{\mu \nu \rho_1 \cdots 
     \rho_n} (\tilde{H}^{2k-1})_{\mu \nu }
      \partial_{\rho_1}(\delta B_{\rho_2 \cdots \rho_n}) \right\} \\
 &=& \int d^{n+2}x \frac {1}{(n-1)!} \epsilon^{\mu \nu \rho_1 \cdots 
     \rho_n} \partial_{\rho_1} A_{\mu \nu } \delta B_{\rho_2 \cdots
      \rho_n}, 
\end{eqnarray*}
where
\begin{equation}
     A_{\mu \nu }=\displaystyle{\frac 
     {\sum_{k=1}^m (-1)^{k-1} \left( \sum_{j=0}^{m-k} S_{1^j}(z) \right) 
      (\tilde{H}^{2k-1})_{\mu \nu }}
     {\sqrt{\sum_{j=0}^m S_{1^j}(z)}} } \ , 
 \label{eqn:Amu}
\end{equation}
and
\begin{equation}
     z=(z_1, \cdots ,z_m), \quad \quad 
     z_j \equiv \frac{1}{2j}Tr(\tilde{H} \eta )^{2j}
 \label{eqn:zj}
\end{equation}
as in Lemma \ref{lem:det}.
Hence we have the field equation: 
$$ \epsilon^{\mu \nu \rho_1 \cdots \rho_n}
    \partial_{\rho_1} A_{\mu \nu }=0. $$
The following two equations correspond to the Bianchi identity and the Yang-Mills equation in the Maxwell theory.
\begin{equation}
  dH=0, 
 \label{eqn:Bia}
\end{equation}
\begin{equation}
  \epsilon^{\mu \nu \rho_1 \cdots \rho_n}
                                    \partial_{\rho_1} A_{\mu \nu }=0.
 \label{eqn:Yan}
\end{equation}
 \section{Reduction of field equations}

We start from the ansatz:
\begin{equation}
  H \equiv i_X \Omega 
 \label{eqn:Ansa}
\end{equation}
where
 $$
 \begin{array}{l}
  \phi =\phi (x_1,\cdots,x_{n+1}) \\
  X=grad \phi \\
  \Omega =dx_1 \wedge \cdots \wedge dx_{n+1}. \\
 \end{array}
$$
$H$ has the following form;
\begin{eqnarray}
 H &=& \sum_{j=1}^{n+1} (-1)^{j-1} (\partial_j \phi ) dx_1 \wedge \cdots \wedge 
        \check{dx_j} \wedge \cdots \wedge dx_{n+1} \label{eqn:vol}\\
   && \hspace{1cm} H_{\rho_1 \cdots \rho_n}
       =\epsilon_{0 \rho_1 \cdots \rho_n \rho_{n+1}} 
         \partial_{\rho_{n+1}} \phi . \nonumber
\end{eqnarray}
Under this ansatz, the Bianchi identity (\ref{eqn:Bia}) becomes 
\begin{equation}
 \Delta \phi =0, 
 \label{eqn:rBia} 
\end{equation}
where $\Delta $ is the $(n+1)$-dimensional Laplacian.

Next, we shall see how our Yang-Mills equation (\ref{eqn:Yan}) becomes under the ansatz.  
The dual form $\tilde{H}$ of $H$ is given by
$$
\left\{
 \begin{array}{rl}
   \tilde{H}_{0 j}=-\partial_j \phi  & \quad (j=1,\cdots,n+1) \\
   \tilde{H}_{\mu \nu}=0 & \quad \mbox{otherwise.}
 \end{array}
\right.
$$
Therefore, the matrix $\tilde{H} \eta $ becomes 
\begin{equation}
 \tilde{H} \eta=
    \left(
     \begin{array}{c|ccc}
       0 & & {}^t \mbox{\boldmath $a$} & \\ \hline
         & & & \\
       \mbox{\boldmath $a$} & & \mbox{\boldmath $0$} & \\
         & & & 
     \end{array}
    \right)
\hspace{10mm}
\mbox{where} \quad 
  \mbox{\boldmath $a$}=-
    \left(
     \begin{array}{c}
      \partial_1 \phi  \\
      \vdots \\
      \partial_{n+1} \phi  
     \end{array}
    \right).
 \label{eqn:Htilde} 
\end{equation}
Putting (\ref{eqn:Htilde}) in (\ref{eqn:zj}) we have 
\begin{equation}
 z_j = \frac{1}{2j}Tr(\tilde{H} \eta )^{2j} = \frac 1j z_1^j \quad (j \geq 2), 
  \quad \quad
 z_1=\frac 12 Tr(\tilde{H} \eta)^2=|grad \phi |^2 
\end{equation}
where $n=2m-1$ or $2m-2$, because 
$z_j$ for $j \geq 2$ is obtained by the following formula:
\begin{equation}
   (\tilde{H}^{2k-1})_{\mu \nu }=z_1^{k-1} \tilde{H}_{\mu \nu } \quad
    (k=1,2,\cdots,m).
 \label{eqn:Hk} 
\end{equation}
We have from (\ref{eqn:Schur})
\begin{equation}
 S_{1^j}(z)=0 \hspace{10mm} (j \geq 2). 
 \label{eqn:Shu}
\end{equation}
By using (\ref{eqn:Hk}) and (\ref{eqn:Shu}), we have from (\ref{eqn:Amu})
 \begin{eqnarray*}
   A_{\mu \nu } &=& 
    \frac{\sum_{k=1}^{m-1} (-1)^{k-1}(1+z_1)(\tilde{H}^{2k-1})_{\mu \nu }
          +(-1)^{m-1}(\tilde{H}^{2m-1})_{\mu \nu }}
         {\sqrt{1+z_1}} \\
    &=& \frac{\tilde{H}_{\mu \nu }}{\sqrt{1+z_1}}.
 \end{eqnarray*}
Here non-trivial components are only
$$ A_{0 j}=-A_{j 0}=-\frac{\partial_j \phi }{\sqrt{1+|grad \phi |^2}} 
   \hspace{10mm}
    (j=1,2,\ldots,n+1). $$
We find that the field equation (\ref{eqn:Yan}) holds if and only if 
\begin{equation}
 \left\{ \phi , |grad \phi |^2 \right\}_{j,k}=0 \hspace{10mm} 
     (1 \le j < k \le n+1) 
 \label{eqn:rYan}
\end{equation}
where $ \quad 
           \{ f,g \}_{j,k} \equiv (\partial_j f)(\partial_k g)-
                            (\partial_k f)(\partial_j g). $ \\
Thus, under the ansatz (\ref{eqn:Ansa}) we have reduced the field equations (\ref{eqn:Bia}) and (\ref{eqn:Yan}) to the following forms: 
$$
 \left\{
    \begin{array}{lcl}
      \mbox{Bianchi identity} \quad (\ref{eqn:Bia}) & \longrightarrow & 
       \Delta \phi =0 \quad (\ref{eqn:rBia}) \\
      \mbox{Yang-Mills equation} \quad (\ref{eqn:Yan}) & \longrightarrow & 
       \left\{ \phi , |grad \phi |^2 \right\}_{j,k}=0 \quad (\ref{eqn:rYan}).
    \end{array}
 \right.
$$
 \section{Soliton solution}

Now we suppose that $ \phi $ has the following form:
$$ 
\phi =\phi (r), \quad  r=
     \sqrt{\displaystyle{\sum_{j=1}^{n+1}}x_j^2}. $$ 
Then (\ref{eqn:rYan}) holds. Moreover, (\ref{eqn:rBia}) holds if and only if 
\begin{equation}
 \phi (r)= c_1 \frac{1}{r^{n-1}}+c_2 
 \label{eqn:1sol}
\end{equation}
where $c_1$ and $c_2$ are constants. Therefore (\ref{eqn:1sol}) is considered as a 1-soliton. \\
For simplicity, we put
$$ c_1=-\frac{1}{n-1} , \quad c_2=0. $$
Then (\ref{eqn:vol}) becomes 
$$ H=\sum_{j=1}^{n+1} (-1)^{j-1} \frac{x_j}{r^{n+1}} dx_1 \wedge \cdots \wedge 
      \check{dx_j} \wedge \cdots \wedge dx_{n+1}. $$
We have $ dH=0 $ on $ {\bf R}^{n+1}-\{ 0 \} $, hence 
there exists a $(n-1)$-form $B$ locally satisfying $ H=dB $, with $B$ being a solution of the field equation(\ref{eqn:Yan}).
Such $(n-1)$-form was constructed by Fujii \cite{Fuj}.
\begin{lem}[Fujii]
 $$
 B = f(r,x_{n+1}) \sum_{j=1}^n (-1)^{n+j} x_j dx_1 
      \wedge \cdots \wedge \check{dx_j} \wedge \cdots \wedge dx_n ,$$
 where
 \begin{equation}
  f(r,x_{n+1}) \equiv \int_0^1 \frac {2^n t^{n-1}}
                    {\{ r-x_{n+1}+t^2 (r+x_{n+1}) \}^n}dt. 
 \end{equation}
\end{lem}
For simplicity, we put
\begin{equation} 
\left\{
 \begin{array}{c}
  a=r-x_{n+1} \\
  b=r+x_{n+1}
 \end{array}
\right.
\label{eqn:ab}
\end{equation}
and
\begin{equation}
f(r,x_{n+1})=\int_0^1 \frac {2^n t^{n-1}}
                  {( a+t^2 b )^n}dt. 
  \label{eqn:int}
\end{equation}
To calculate $f(r,x_{n+1})$, he used in \cite{Fuj} the following method. \\
If \ $n=2m-1$,
\begin{eqnarray}
   f(r,x_{n+1}) 
    &=& \frac {2^{2m-1}}{(2m-2)!} 
         \left( \frac {\partial }{\partial a} \right)^{m-1}
         \left( \frac {\partial }{\partial b} \right)^{m-1}
          \int_0^1 \frac {1}{a+t^2 b}dt \nonumber\\
    &=& \frac {2^{2m-1}}{(2m-2)!} 
         \left( \frac {\partial }{\partial a} \right)^{m-1}
         \left( \frac {\partial }{\partial b} \right)^{m-1}
          \left( \frac {\arctan \sqrt{\frac ba}}{\sqrt{ab}} \right). 
 \label{eqn:odd}
\end{eqnarray}
If \ $n=2m-2$,
\begin{eqnarray}
   f(r,x_{n+1}) 
    &=& \frac {-2^{2m-2}}{(2m-3)!} 
         \left( \frac {\partial }{\partial a} \right)^{m-1}
         \left( \frac {\partial }{\partial b} \right)^{m-2}
          \int_0^1 \frac {t}{a+t^2 b}dt \nonumber\\
    &=& \frac {2^{2m-2}}{(2m-3)!} 
         \left( \frac {\partial }{\partial a} \right)^{m-2}
         \left( \frac {\partial }{\partial b} \right)^{m-2}
         \left( \frac {1}{2a(a+b)} \right).
 \label{eqn:even}
\end{eqnarray}
Let us calculate (\ref{eqn:odd}) and (\ref{eqn:even}) explicitly. Putting
$$ f_n \equiv f(r,x_{n+1}) $$
and 
$$
X_n \equiv 
 \left\{
  \begin{array}{cc}
   \left( \displaystyle{\frac {\partial }{\partial a}} \right)^{m-1}
   \left( \displaystyle{\frac {\partial }{\partial b}} \right)^{m-1} X_1
    & \hspace{1cm} \mbox{if} \ n=2m-1 \\
   \left( \displaystyle{\frac {\partial }{\partial a}} \right)^{m-2}
   \left( \displaystyle{\frac {\partial }{\partial b}} \right)^{m-2} X_2
    & \hspace{1cm} \mbox{if} \ n=2m-2 
  \end{array}
 \right.
$$
where
\begin{equation}
  X_1=\frac {\arctan \sqrt{\frac ba}}{\sqrt{ab}} \ , \quad 
  X_2=\frac {1}{2a(a+b)}, 
 \label{eqn:ini}
\end{equation}
we have by (\ref{eqn:odd}) and (\ref{eqn:even}), 
\begin{equation}
  f_n=\frac {2^n}{(n-1)!} X_n.
 \label{eqn:Xn}
\end{equation}
Therefore we have only to determine $X_n$.
\begin{prop}\label{prop:cart}
 If \ $n=2m-1$,
 \begin{eqnarray*}
 X_n &=& \frac {(2m-2)!(2m-3)!!}{(2m-2)!!(2^2 ab)^{m-1}}
          \left\{ 
           \frac {\arctan \sqrt{\frac ba}}{\sqrt{ab}} \right. \\
     && \hspace{3cm} \left. +\sum_{k=1}^{m-1} 
             \frac{(2k-2)!!(b-a)(2^2 ab)^{k-1}}{(2k-1)!!(a+b)^{2k}}
          \right\}. 
 \end{eqnarray*}
 If \ $n=2m-2$,
 \begin{eqnarray*}
 X_n &=& \frac {(2m-3)!(2m-4)!!}{(2m-3)!!(2^2 ab)^{m-2}}
          \left\{ 
           \frac {1}{2a(a+b)} \right. \\
    && \hspace{3cm} \left. + \sum_{k=2}^{m-1} 
             \frac{(2k-3)!!(b-a)(2^2 ab)^{k-2}}{(2k-2)!!(a+b)^{2k-1}}
          \right\}. 
 \end{eqnarray*}
Hence by (\ref{eqn:ab}) and (\ref{eqn:Xn}),
we obtain the following:
 if \ $n=2m-1$, 
  $$
   f(r,x_{n+1})=
        \frac {(2m-3)!!}{(2m-2)!!(r^2-x_{2m}^2)^{m-1}}   
         \left\{ 
           \frac{2\arctan \sqrt{\frac{r+x_{2m}}{r-x_{2m}}}}
                {(r^2-x_{2m}^2)^{\frac12}} \right.  $$
  $$ \hspace{7cm}  \left. +\sum_{k=1}^{m-1} 
             \frac{(2k-2)!!x_{2m}(r^2-x_{2m}^2)^{k-1}}{(2k-1)!!r^{2k}}
          \right\} ,
  $$
 if \ $n=2m-2$, 
  $$ \hspace{-3cm} 
   f(r,x_{n+1})=
        \frac {(2m-4)!!}{(2m-3)!!(r^2-x_{2m-1}^2)^{m-1}} $$
  $$ \hspace{5cm} \times \left\{ 
            1+\sum_{k=1}^{m-1} 
             \frac{(2k-3)!!x_{2m-1}(r^2-x_{2m-1}^2)^{k-1}}{(2k-2)!!r^{2k-1}}
          \right\} .$$
\end{prop}
{\it Proof}: By calculating $dB=H$, we have differential equations for $f_n=f(r,x_{n+1})$ 
\begin{equation}
  \sum_{j=1}^{n+1} x_j \partial_j f_n = -n f_n,
 \label{eqn:dif1}
\end{equation}
\begin{equation}
  \partial_{n+1} f_n = \frac {1}{r^{n+1}}.
 \label{eqn:dif2}
\end{equation}
We change the variables
$$ 
\left\{
 \begin{array}{c}
  a=r-x_{n+1} \\
  b=r+x_{n+1}.
 \end{array}
\right.
$$
Then (\ref{eqn:dif1}) and (\ref{eqn:dif2}) become
\begin{equation}
  S f_n = -n f_n,
 \label{eqn:difS}
\end{equation}
\begin{equation}
  T f_n = \frac {2^n}{(a+b)^n}.
 \label{eqn:difT}
\end{equation}
where
$$ S=a \frac {\partial }{\partial a}+b \frac {\partial }{\partial b} \ ,
  \quad
   T=-a \frac {\partial }{\partial a}+b \frac {\partial }{\partial b}.
$$
By (\ref{eqn:Xn}), (\ref{eqn:difS}) and (\ref{eqn:difT}), we have 
$$
S X_n = -n X_n,
$$
$$
T X_n = \frac {(n-1)!}{(a+b)^n}.
$$
Noting
$$
X_n=\left( \frac {\partial }{\partial a} \right)
    \left( \frac {\partial }{\partial b} \right) X_{n-2} \ , \quad 
\left( \frac {\partial }{\partial a} \right)
\left( \frac {\partial }{\partial b} \right)
 =\frac {1}{2^2 ab}(S^2-T^2)
$$
and 
$$
T^2 X_n=(n-1)! T \left\{ \frac {1}{(a+b)^n} \right\} 
       =-n! \frac {b-a}{(a+b)^{n+1}},
$$
we have
\begin{eqnarray}
 X_n &=& \left( \frac {\partial }{\partial a} \right)
         \left( \frac {\partial }{\partial b} \right) X_{n-2} \nonumber\\
     &=& \frac {1}{2^2 ab}(S^2-T^2) X_{n-2} \nonumber\\
     &=& \frac {1}{2^2 ab}
          \left\{ 
           (n-2)^2 X_{n-2}+(n-2)! \frac {b-a}{(a+b)^{n-1}} 
          \right\}. \label{eqn:rec}
\end{eqnarray}
By using the recursion formula (\ref{eqn:rec}) and initial conditions (\ref{eqn:ini}), we finally obtain our results. \quad \qed

\begin{rmk}
We give another calculation of $f(r,x_{n+1})$ in the appendix. 
\end{rmk}

Therefore we have 
$$ 
B_{\rho_1 \cdots \rho_{n-1}}=
 \left\{
  \begin{array}{ll}
    -f(r,x_{n+1}) 
      \epsilon_{\rho_1 \cdots \rho_{n-1} \rho_n} x_{\rho_n}
      & \quad \mbox{if} \quad \rho_1, \cdots ,\rho_{n-1}=1,\cdots,n \\
    0 & \quad \mbox{otherwise}
  \end{array}
 \right.
$$ 
where \ $\epsilon_{1 2 \cdots n}=-1$.

\begin{prop} In polar coordinates
$$
\left\{
 \begin{array}{l}
  x_1=r \sin \theta_1 \cos \theta_2 \\
  \vdots \\
  x_{n-1}=r \sin \theta_1 \sin \theta_2 \cdots \sin \theta_{n-1} 
               \cos \theta_n \\ 
  x_n=r \sin \theta_1 \sin \theta_2 \cdots \sin \theta_{n-1} 
               \sin \theta_n \\ 
  x_{n+1}=r \cos \theta_1,
 \end{array}
\right.
$$
then 
$$
\left\{
 \begin{array}{l}
  H=(-1)^n \sin^{n-1} \theta_1 \sin^{n-2} \theta_2 \cdots \sin \theta_{n-1}
      d \theta_1 \wedge \cdots \wedge d \theta_n \\
  B=(-1)^n g(\theta_1) \sin^{n-2} \theta_2 \cdots \sin \theta_{n-1}
      d \theta_2 \wedge \cdots \wedge d \theta_n
 \end{array}
\right.
$$
where

 if \ $n=2m-1$,
$$ g(\theta_1)=
           -\frac {(2m-3)!!}{(2m-2)!!} 
                \left(
                 \pi -\theta_1+\sum_{k=1}^{m-1} \frac {(2k-2)!!}{(2k-1)!!}
                  \cos \theta_1 \sin^{2k-1} \theta_1
                \right) ,$$

 if \ $n=2m-2$,
$$ g(\theta_1)=
           -\frac {(2m-4)!!}{(2m-3)!!} 
                \left(
                 1+\sum_{k=1}^{m-1} \frac {(2k-3)!!}{(2k-2)!!}
                  \cos \theta_1 \sin^{2k-2} \theta_1
                \right) .$$
\end{prop}
 \section{Discussion}

We have defined the Born-Infeld action in $(n+2)$-dimensional Minkowski space and have constructed a solution of the field equation. 

The 1-soliton (\ref{eqn:1sol}) is a solution of (\ref{eqn:rBia}) and (\ref{eqn:rYan}). Therefore, it is natural to ask whether $N$-solitons $(N \geq 2)$ 
\begin{equation}
 c_0+\sum_{j=1}^N \frac {c_j}{|x-p_j|^{n-1}} 
 \label{eqn:Nsol}
\end{equation} 
are solutions of (\ref{eqn:rBia}) and (\ref{eqn:rYan}) or not. But (\ref{eqn:Nsol}) don't satisfy (\ref{eqn:rYan}) though (\ref{eqn:Nsol}) satisfy (\ref{eqn:rBia}).

Another problem is to understand the geometrical meaning of the gauge field $B$.  
\\

{\it Acknowledgment}\\
The author is very grateful to Kazuyuki Fujii for helpful suggestions and comments and to Tosiaki Kori and Michitomo Nishizawa for valuable discussions.

\appendix
\section{Appendix}
We give another calculation of the integral (\ref{eqn:int})
$$
f(r,x_{n+1})=\int_0^1 \frac {2^n t^{n-1}}
                  {( a+t^2 b )^n}dt. $$
First, we have 
$$
f(r,x_{n+1})=\frac{1}{(ab)^{\frac n2}} 
             \int_0^1 \frac{2^n}
                 {\left\{ 
                   (\sqrt{\frac ba}t)^{-1}+(\sqrt{\frac ba}t)
                  \right\} ^n}
             \frac{dt}{t}.
$$
Here we change the variable $t$ as
$$
t=e^{-s}
$$
and put for simplicity 
\begin{equation}
  e^{\alpha}=\sqrt{\frac ba}
 \label{eqn:alpha}
\end{equation}
to obtain
\begin{eqnarray}
   f(r,x_{n+1}) 
    &=& \frac{1}{(ab)^{\frac n2}} 
          \int_0^{\infty} \frac{2^n}
                 {\left\{ 
                    e^{s-\alpha }+e^{-(s-\alpha )}
                  \right\} ^n} ds \nonumber\\
    &=& \frac{1}{(ab)^{\frac n2}} 
          \int_0^{\infty} \cosh^{-n}(s-\alpha )ds \nonumber\\
    &=& \frac{1}{(ab)^{\frac n2}} I[-n],
         \label{eqn:frx}
\end{eqnarray}
where
$$
I[-n] \equiv \int_0^{\infty} \cosh^{-n}(s-\alpha )ds .
$$
Therefore we have only to calculate $I[-n]$. 

For $n \ne 1$, we have the next formula by using the integration by parts.
\begin{equation}
 I[-n] = \frac{n-2}{n-1}
          \left\{
           I[-(n-2)]+\frac{1}{n-2}\cosh^{-(n-1)}\alpha \ \sinh \alpha
          \right\} .
  \label{eqn:part}
\end{equation}
On the other hand, we have by (\ref{eqn:alpha}), 
\begin{equation}
  \cosh \alpha = \frac{a+b}{2\sqrt{ab}} \ , \quad
  \sinh \alpha = \frac{b-a}{2\sqrt{ab}} .
 \label{eqn:hyp}
\end{equation}
Substituting (\ref{eqn:hyp}) into (\ref{eqn:part}), we have 
\begin{equation}
 I[-n] = \frac{n-2}{n-1}
          \left\{
           I[-(n-2)]+\frac{(b-a)(2\sqrt{ab})^{n-2}}
                          {(n-2)(a+b)^{n-1}}
          \right\} .
  \label{eqn:formula}
\end{equation}
By using the recursion formula (\ref{eqn:formula}) and 
$$
I[-1]=2\arctan \sqrt{\frac ba} 
        \hspace{20mm} \mbox{if} \quad n=2m-1, 
$$
$$ \hspace{-1mm}
I[-2]=\frac{2b}{a+b} 
        \hspace{32mm} \mbox{if} \quad n=2m-2, 
$$
we obtain the following: \\
if \ $n=2m-1$, 
\begin{eqnarray*}
I[-n] &=& \frac {(2m-3)!!}{(2m-2)!!}
          \left\{ 
              2\arctan \sqrt{\frac ba} \right. \\
      && \hspace{3cm} \left. +\sum_{k=1}^{m-1} 
             \frac{(2k-2)!!(b-a)(2^2 ab)^{k-\frac 12}}{(2k-1)!!(a+b)^{2k}}
          \right\} ,
\end{eqnarray*}
if \ $n=2m-2$, 
\begin{eqnarray*}
I[-n] &=& \frac {(2m-4)!!}{(2m-3)!!}
          \left\{ 
           \frac {2b}{a+b} \right. \\
      && \hspace{3cm} \left. + \sum_{k=2}^{m-1} 
             \frac{(2k-3)!!(b-a)(2^2 ab)^{k-1}}{(2k-2)!!(a+b)^{2k-1}}
          \right\} .
\end{eqnarray*}
(\ref{eqn:frx}) and the equations above lead to the conclusion. 
If \ $n=2m-1$,
\begin{eqnarray*}
   f(r,x_{n+1}) 
    &=& \frac {(2m-3)!!}{(2m-2)!!(ab)^{m-1}}
          \left\{ 
           \frac {2\arctan \sqrt{\frac ba}}{\sqrt{ab}} \right. \\
    && \hspace{3cm} \left. +2\sum_{k=1}^{m-1} 
             \frac{(2k-2)!!(b-a)(2^2 ab)^{k-1}}{(2k-1)!!(a+b)^{2k}}
          \right\} ,  
\end{eqnarray*}
if \ $n=2m-2$,
\begin{eqnarray*}
   f(r,x_{n+1}) 
    &=& \frac {(2m-4)!!}{(2m-3)!!(ab)^{m-1}}
          \left\{ 
           \frac {2b}{a+b} \right. \\
    && \hspace{3cm} \left. +\sum_{k=2}^{m-1} 
             \frac{(2k-3)!!(b-a)(2^2 ab)^{k-1}}{(2k-2)!!(a+b)^{2k-1}}
          \right\} .
\end{eqnarray*}
 
\end{document}